\documentclass{ws-p8-50x6-00}

\begin{document}

\title{Linear Collider Signals of Anomaly Mediated Supersymmetry
Breaking}

\author{Sourov Roy}

\address{Department of Theoretical Physics, Tata Institute of
Fundamental Research \\ Homi Bhabha Road,  
Mumbai - 400 005, INDIA\\ 
E-mail: sourov@theory.tifr.res.in}

\address{Present address : Department of Physics, Technion,
32000 Haifa, Israel\\ 
E-mail: roy@physics.technion.ac.il}

\maketitle

\abstracts{
Diagnostic signals of the minimal model of anomaly mediated supersymmetry
breaking are discussed in the context of a $\sqrt{s}$ = 1 TeV $e^+e^-$
linear collider.} 

\section{Introduction}
The idea of anomaly mediated supersymmetry breaking 
(AMSB)~\cite{giudice,randall}, where the SUSY breaking is conveyed to the
observable sector by the super-Weyl anomaly, has attracted a lot of 
attention in recent times. This can lead to characteristically distinct 
and unique signatures in various collider experiments~
\cite{feng,wells,fengmoroi}. In this work, we have concentrated on 
the signals of AMSB which can be studied~\cite{dkgprsr,kundudkprsr} in a 
high energy $e^+e^-$ linear collider at a c.m.  energy $\sqrt{s} = 1$ TeV. 
Anomaly mediation is a special case of gravity mediation with no 
tree-level couplings between the superfields of the hidden and the
observable sectors. This is realized, for instance, when the two sectors
are localized on two parallel but distinct 3-branes located in a higher
dimensional bulk. This scenario has several distinct and unique features 
with important phenomenological consequences. The gravitino is rather 
massive ($m_{3/2} \sim$ tens of TeV), the lightest supersymmetric particle 
(LSP, $\widetilde \chi^0_1$) is predominantly a Wino and it is nearly 
mass-degenerate ( mass-splitting $\Delta M$ $<$ 1 GeV) with the lighter 
chargino ($\widetilde \chi^\pm_1$) which is also a near-Wino. Such a 
chargino will be long lived and is likely to leave a heavily ionizing 
charged track in the detector and/or a associated soft pion from the 
decay $\widetilde \chi^\pm_1 \rightarrow \widetilde \chi^0_1 + \pi^\pm$.
The most glaring problem of AMSB is that it predicts negative mass squares 
for the sleptons. In the minimal AMSB model this problem is solved 
by adding a universal constant term $m^2_0$ to the expressions for squared 
scalar masses. The set of parameters which define the minimal AMSB model is
$\{m_{3/2},m_0,sign(\mu)$, $\tan\beta$\}. The AMSB sparticle mass spectrum 
can be broadly classified into two categories:  

\begin{itemize}
\item Spectrum A: $\widetilde \chi^0_1(\approx \widetilde \chi^\pm_1)
< \tilde \nu < {\tilde e}_R (\approx {\tilde e}_L) < \widetilde \chi^0_2$
\item Spectrum B: $\widetilde \chi^0_1(\approx \widetilde \chi^\pm_1)
< \widetilde \chi^0_2 < \tilde \nu < {\tilde e}_R (\approx {\tilde e}_L).$
\end{itemize}
We discuss our signals for them separately.

\section{Results}
A list of all possible final states for both the spectra mentioned 
above, is given in Table~1 for one pion channels. The signals are always 
accompanied by missing energy. A similar list for two pion channels is given 
in Ref.~\cite{kundudkprsr}. Cross sections for the various production
processes have been calculated for two values of $\tan\beta$, namely 10 
and 30 and for $\mu > 0$. In most of the allowed parameter space the 
signal cross sections are $\sim$ 10-100 fb. For example, in the 
worst cases of the signal cross sections, assuming an integrated luminosity 
of $500~{\rm fb^{-1}}$, one would expect 13165 signal events from spectrum 
A, while 2910 signal events are predicted from spectrum B for the signal
$e^\pm + \pi^\mp + E_T \hspace{-1.1em}/\;\: $. The kinematic distributions 
of the final state particles for the same signal have been studied for a 
sample point in the AMSB parameter space corresponding to Spectrum A. They 
are shown in Fig. 1. The chargino decay length distribution shows that a 
substantial number of events do have a large decay length so that the 
charged track can be seen. The signals analyzed here are essentially free 
of Standard Model background. 

\begin{table}[t]
\caption{Possible one or multilepton signal with one soft pion.
\label{tab:singlepi}}
\begin{center}
\footnotesize
\begin{tabular}{|c|c|c|} \hline
{\bf Spectrum} & {\bf Signals} & {\bf Parent Channels} \\ \hline
 & $e~\pi$ & $\tilde \nu \bar {\tilde \nu},~~{\tilde e}_L {\tilde e}_L,
~~{\tilde e}_L
{\tilde e}_R,~~\widetilde \chi_1^0 \widetilde \chi_2^0,~~\widetilde
\chi_2^0 \widetilde \chi_2^0$ \\
 & $\mu~\pi$ & $\tilde \nu \bar {\tilde \nu},~~\widetilde \chi_1^0 \widetilde
\chi_2^0,~~ \widetilde \chi_2^0 \widetilde \chi_2^0$ \\
{\bf A} & $e~e~\ell~\pi$ & ${\tilde e}_R {\tilde e}_R,~~{\tilde e}_L
{\tilde e}_R,~~\widetilde \chi_1^0
\widetilde \chi_2^0,~~
\widetilde \chi_2^0 \widetilde \chi_2^0$ \\
 & $\mu~\mu~\ell~\pi$ & $\widetilde \chi_1^0 \widetilde \chi_2^0,~~
\widetilde \chi_2^0 \widetilde \chi_2^0$ \\
 & $\ell_1~\ell_1~\ell_2~\ell_2~\ell_3~\pi$ & $\widetilde \chi_2^0
\widetilde \chi_2^0$
($\ell_{1,2,3}=e,~\mu$) \\ \hline
 & $e~\pi$ & $\tilde \nu \bar {\tilde \nu},~~{\tilde e}_L {\tilde e}_L,~~
{\tilde e}_L {\tilde e}_R,~~
\widetilde \chi_1^0 \widetilde \chi_2^0,~~\widetilde \chi_2^0 \widetilde
\chi_2^0$ \\
 & $\mu~\pi$ & $\tilde \nu \bar {\tilde \nu},~~{\tilde e}_L {\tilde
e}_L,~~\widetilde \chi_1^0 \widetilde \chi_2^0,~~
\widetilde \chi_2^0 \widetilde \chi_2^0$ \\{\bf B} & $e~\ell_1~\ell_2~\pi$ & 
${\tilde e}_R {\tilde e}_R,~~{\tilde e}_L {\tilde e}_R,~~
  {\tilde e}_L {\tilde e}_L,~~ \tilde \nu \bar {\tilde \nu},~~
\widetilde \chi_2^0 \widetilde \chi_2^0$ \\
 & $\mu~\mu~\mu~\pi$ & $\widetilde \chi_2^0 \widetilde \chi_2^0,~~
\tilde \nu \bar {\tilde \nu}$ \\
 & $e~e~\ell_1~\ell_1~\ell_2~\pi$ & ${\tilde e}_L {\tilde e}_L,~~{\tilde
e}_R {\tilde e}_R, ~~{\tilde e}_L {\tilde e}_R$ ($\ell_{1,2}=e,~\mu$)\\ \hline
\end{tabular}
\end{center}
\end{table}

\begin{figure}[t]
\begin{center}
\epsfxsize=12pc
\epsfbox{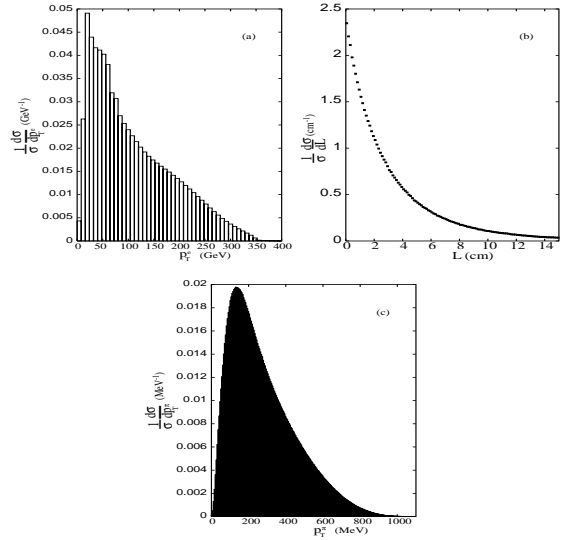}
\caption{Normalized kinematic distributions of decay products:
$(a)$ $p_T$ of charged lepton,
$(b)$ decay length of the lighter chargino,
and $(c)$ $p_T$ of the charged pion arising from
$e^\pm + \pi^\mp + E_T \hspace{-1.1em}/\;\:$ signal for spectrum A.
The AMSB input parameters are $m_{3/2}=44$~TeV, $\tan\beta = 30$,
$\mu > 0$ and $m_0 = 410$ GeV.
\label{fig:kin_dist}}
\end{center}
\end{figure}

\section{Conclusion}
We have investigated possible signals of minimal AMSB model
in a 1 TeV c.m energy $e^+e^-$ linear collider. The model 
is characterized by nearly degenerate lightest neutralino 
($\widetilde \chi^0_1$) and the lighter chargino ($\widetilde
\chi^\pm_1$), resulting in a heavily ionizing charged track and/or a
detectable soft $\pi^\pm$ from the long-lived decay $\widetilde \chi^\pm_1 
\rightarrow \widetilde \chi^0_1$ + soft $\pi^\pm$. The generated events 
triggered by fast charged leptons are large in number and the allowed 
region of the parameter space can be comprehensively probed.   

\section*{Acknowledgments}
I thank the organisers of SUSY'01 at JINR, Dubna, Russia for the
hospitality.

\end{document}